\documentclass[11pt]{article}
\usepackage{geometry}
\usepackage[utf8]{inputenc}
\usepackage{natbib}
\usepackage{lscape}
\usepackage{amsmath}
\usepackage{amsfonts}
\usepackage{amssymb}
\usepackage{graphicx}
\usepackage{algorithm}
\usepackage{float}
\usepackage{algpseudocode}
\usepackage{fancyhdr}
\usepackage{subcaption}
\usepackage{url}

\usepackage{rotating}

\usepackage{multirow}
\usepackage[table]{xcolor}

\usepackage[none]{hyphenat}

\begin{document}

\title{\LARGE \textbf{
Uncertainty Quantification of State Variables Trajectories in the Context of Inverse Problems: An Approach from Bayesian Inference and FDA}\\ 
\vspace{1cm} }

\sloppy

\author{Luis Alejandro Baena-Marín$^\dag$,
Juan Daniel Molina$^\ddag$\footnote{Corresponding author: Faculty of Economic and Administrative Sciences, Institución Universitaria ITM, Calle 73 No. 76A – 354, Medellín, 50034, Antioquia, Colombia. E-mail: juanmolina1192@correo.itm.edu.co}, \
Juan Camilo Bermúdez-Colorado$^\dag$\\ 
and Nicolás Moreno$^\dag$\\ 
\vspace{0.2cm}\\ 
$^\dag$School of Applied Sciences and Engineering, Universidad EAFIT;\\
$^\ddag$Faculty of Economic and Administrative Sciences, Institución Universitaria ITM\\
}

\date{ }

\maketitle

\begin{abstract}

In this article, we address the problem of uncertainty quantification of state variables in the context of inverse problems. Inverse problems are associated with phenomena that can be represented through ordinary or partial differential equations, for which observations or data are available, but the values of the parameters that characterize the equations, the initial or boundary conditions are not known. Currently, the literature offers very few alternatives for analyze the propagation of uncertainty of state variables, which are limited to constructing pseudo-credible regions or quantifying their uncertainty at isolated points. We propose a methodology that combines tools of Bayesian inference, with Hamiltonian Monte Carlo sampling employed for efficient posterior exploration,  and functional data analysis, specifically the Modified Band Depth method, to determine credible regions for the trajectories of the state variables throughout the time horizon of interest. We expose a methodology validation  through a simulation study, which shows that our proposal captures a higher proportion of state variable trajectories than traditional pointwise analysis methods, our proposal generated credible regions that contained the true trajectory of the state variables 96.4\% of the times, versus 80\% that the regions of the pointwise method did. Furthermore, we demonstrate its application to a non-trivial model associated with a neurological phenomenon, for which the methodology effectively captures the time-dependent dynamics of the state variables.

\hfill 

\end{abstract}

\section{Introduction}

Uncertainty quantification (UQ) for inverse problems is an area of knowledge that has gained prominence in recent years. It is associated with probabilistic and inferential analysis processes of the propagation of uncertainty of parameters and variables within the context of phenomena that can be represented by differential equations (DE), ordinary differential equations (ODE), or partial differential equations (PDE). These processes typically require multidisciplinary theoretical and practical tools from fields such as Statistics, Applied Mathematics, and Computer Science, as well as from the specific field of knowledge from which the phenomenon of interest originates.\\

In recent years, the use of the Bayesian statistical approach to perform UQ for inverse problems has also expanded. This methodology is known as Bayesian uncertainty quantification (BUQ). Some reviews of this topic can be found in \cite{kaipio2006}, \cite{fox2013} and \cite{Dashti2017}. Recent applications of BUQ have been very varied, in areas such as image processing \citep{giovannelli2015,chama2012}, geothermal energy \citep{cui2011,cui2019}, ecology \citep{hutchinson2017}, heat transfer \citep{kaipio2011}, tumor growth \citep{collis2017,kahle2019}, among others.\\  

To fix ideas, we consider a phenomenon represented by means of an ODE system. Assuming that we have observations $Y=(y_1, \dots , y_n)$ taken at times $t=(t_1, \dots , t_n)$, with the following structure
\begin{equation}
	y_i = X_{\theta}(t_i)  + \varepsilon_i, \; \; \text{for} \; \; i = 1, \cdots n.
\end{equation}
Where $\varepsilon_i$ is the random noise associated with the observation $y_i$ (it is common, for example, to have $\varepsilon_i \sim \mathcal{N }(0 \, , \, \sigma^2)$), and $X_\theta \in \mathbb{R}^{d}$ represents the solution of the following ODE system,
\begin{equation} \label{PVI}
	\frac{d X_\theta}{dt} = F(X_\theta , t, \theta), \; \; \; X_\theta (t_0) = X_0,
\end{equation}
$\theta \in \mathbb{R}^{p}$ and $F:\mathbb{R}^{d}\times \mathbb{R}^{+} \times \mathbb{R}^{p} \rightarrow \mathbb{R}^{d}$ is a known function. The expression \eqref{PVI} is also known as an initial value problem. Finding $X_\theta$ knowing the true value of $\theta$ and the initial conditions is called the direct problem or the forward map, i.e., $\theta \mapsto X_\theta$. However, if the goal is to make an inference about $\theta$ from observations $Y$, we speak of an inverse problem.\\

In the case where each observation error represents a gaussian additive noise, the likelihood function $f(\, y \, \lvert \, \theta \,)$ is
$$
f(\, y \, \lvert \, \theta \,) = \left( 2\pi \sigma^2 \right)^{-\frac{n}{2}}  \mathrm{exp}\left[ - \frac{1}{2\sigma^2} \sum_{i=1}^{n} \left( \, y_i - X_{\theta} \left(  t_i \right) \, \right) ^2  \right]. 
$$

Thus, each time we evaluate $f(\, y \, \lvert \, \theta \,)$ we must solve $X_{\theta}$. Considering $\pi(\theta)$ as the prior distribution of $\theta$, the posterior distribution of $\theta$, $\pi(\, \theta \, \lvert \, y \,)$, according to the Bayes' theorem, is
$$
\pi(\, \theta \, \lvert \, y \,) = \frac{f(y \, \lvert \, \theta) \pi(\theta)}{\int f(y \, \lvert \, \phi) \pi(\phi) d\phi} = \frac{f(y \, \lvert \, \theta) \, \pi(\theta)}{z(y)},
$$ 
where $z(y) = \int f(y \, \lvert \, \phi) \pi(\phi) d\phi$, is known as the normalization constant of the posterior distribution.\\

The specific problem considered in this article is to analyze the propagation of uncertainty of the vector of state variables $X_\theta$ throughout the time horizon of interest, using functional data analysis (FDA) tools. Currently, the literature offers very few alternatives for this problem. One option considered by some authors to establish pseudo-credible regions for $X_\theta$, is to generate the respective solutions of the forward map from multiple values of $\theta$, obtained from its posterior distribution via MCMC, and finally plot all the different trajectories of the state variables.\\

 A methodology for performing UQ for $X_\theta$ in the context of inverse problems, was proposed by \cite{Molina2025}. This methodology involves generating credible regions for each point in time, $t_i$, based on the values of $\theta$, obtained from its posterior distribution. However, this methodology does not offer a coherent uncertainty analysis over the entire time horizon. Another possible option is to consider the posterior predictive distribution of $X_\theta$. But this, again, would only generate credible regions valid for isolated points in time.\\

FDA has come up as a powerful tool for dealing with data that are functional, such as curves. One notable approach for assessing the centrality and variability of functional data is the Modified Band Depth (MBD) method proposed by \cite{LopezPintado2009}. In essence, MBD evaluates how frequently a given function remains within the envelope defined by all pairs of sample curves, thereby providing a ranking of curves according to their depth or centrality in the dataset. This ranking enables the identification of a functional median and the construction of functional boxplots, which in turn can be adapted to define credible regions for entire trajectories.\\

Recent FDA developments include methods for constructing data-depth (DD) plots that serve as homogeneity tests for functional data. For instance, \cite{CALLESALDARRIAGA2021104420} proposed a DD-plot–based homogeneity test which visually compares functional distributions by assessing whether DD-plots concentrate along the diagonal. Similarly, \cite{lesmes2024eeclassifier} introduced an extremality-based classifier using modified epigraph and hypograph indexes to capture the shape and variability of temporal data. These FDA approaches enable the construction of simultaneous credible regions for entire functional trajectories, thereby preserving the entire temporal coherence. \\

The objective of this article is to use tools of FDA and Bayesian inference to construct a methodology for UQ of the vector of state variables $X_\theta$, in the context of inverse problems, that capture, in a coherent way the temporal dynamics. From determining credible regions for the trajectories of the state variables throughout the time horizon of interest.\\

The structure and organization of this article are as follows: In Section 2, we set out the details of the proposed methodology, in section 3 we develop a validation of the methodology through a simulation study, in section 4 we present an application in a non-trivial context associated with neurological models, and finally, in section 5 the conclusions associated with this article are raised.

\section{Methodology propose to make Uncertainty Quantification for Inverse Problems}

In this section, we describe the conceptual elements on which the methodology we propose for uncertainty quantification of state variables in the context of inverse problems is based. We also present the details of the methodology.

\subsection{Bayesian Uncertainty Quantification}

From the Bayesian statistical approach, the parameter vector, $\theta$, is considered as a random variable, and its uncertainty is measured at two times: prior to experimentation or data collection and after it. Therefore, we speak of prior and posterior distributions for $\theta$. The prior distribution, $\pi (\theta)$, can be based on background
knowledge reported in the literature regarding the phenomenon, or previously available data \citep{gelman}. Non-informative prior distribution can also be chose if
there is not much previous information about $\theta$, or if greater objectivity is sought \citep{berger2006}. The posterior distribution of $\theta$, $\pi(\, \theta \, \lvert \, y \,)$ is calculated via Bayes' theorem, as we shown previously. However, in most cases it is not possible to have an analytic solution of $\pi(\, \theta \, \lvert \, y \,)$. The strategy followed is to use methods of the Markov Chain Monte Carlo (MCMC) type that allow generating values of the posterior distribution of $\theta$.\\

MCMC methods represent a set of algorithms that allow obtaining samples from a given objective probability distribution, from which it is difficult to sample directly. These methods are based on building a Markov chain whose equilibrium distribution is the target distribution. Thus, the states of the Markov chain after it has reached stationary state represent samples from the target distribution. The MCMC method used in this article is the Hamiltonian Monte Carlo (HMC) \citep{neal2011}, which introduces auxiliary momentum variables and leverages the geometry of the parameter space to make efficient proposals.\\ 

For a parameter vector $\theta$ and its corresponding momentum vector $p$, the Hamiltonian is $H(\theta, p) = -\log[\, \pi(\theta \, \lvert \, Y) \,] + \frac{1}{2}p^TM^{-1}p$, where $M$ is a mass matrix that we adapt during sampling. The algorithm proceeds as follows,

\begin{algorithm}[H]
\caption{Hamiltonian Monte Carlo}
\begin{algorithmic}[1]
\State \textbf{Input:} Initial position $\theta^{(0)}$, step size $\epsilon$, path length $L$
\State \textbf{Output:} Samples $\{\theta^{(1)}, \ldots, \theta^{(T)}\}$
\While {$t < T$}
   \State Sample momentum $p \sim \mathcal{N}(0, M)$
   \State $(\theta^*, p^*) \gets (\theta^{(t-1)}, p)$
   \State Leapfrog integration:
   \State \quad $p^* \gets p^* + \frac{\epsilon}{2}\nabla_\theta \log[\, \pi(\theta^* \, \lvert \,Y \,]$
   \State \quad $\theta^* \gets \theta^* + \epsilon M^{-1}p^*$
   \State \quad $p^* \gets p^* + \frac{\epsilon}{2}\nabla_\theta \log[\, \pi(\theta^* \, \lvert \,Y) \,]$
   \State $u \sim \text{Uniform}(0,1)$
   \State $\alpha \gets \min\{1, \exp(H(\theta^{(t-1)}, p) - H(\theta^*, p^*))\}$
   \If{$u < \alpha$}
       \State $\theta^{(t)} \gets \theta^*$
   \Else
       \State $\theta^{(t)} \gets \theta^{(t-1)}$
   \EndIf
\EndWhile
\end{algorithmic}
\end{algorithm}

In practice, we implement a variant of HMC through the Python package PyMC3, the No U-Turn Sampler (NUTS) \citep{hoffman2014}, which automatically selects the path length $L$ and adapts the step size $\epsilon$ and mass matrix $M$. 

\subsubsection{BUQ of state variables}
As we mentioned previously, \cite{Molina2025} propose a methodology for performing UQ for the vector of state variables, $X_\theta$, in the context of inverse problems. This methodology involves generating credible regions for each point in time $t^*$ within the analysis period, based on the values of $\theta$, obtained from its posterior distribution. Thus, the methodology they proposed consists of the following algorithm:

	\begin{algorithm}[H]
			\caption{BUQ of State Variables}
			\begin{algorithmic}[1]
				\Require \text{Data} $Y=(y_1, \dots , y_n)$, 
                \Require \text{right hand side of DE system} $F(\, \cdot \,)$, 
                \Require \text{prior distribution} $\pi(\cdot)$, 
                \Require likelihood function $f(y \, \lvert \, \theta )$   
			\For{$i \gets 1$ to $N$}
				    \State \text{Sample} $\theta_i \sim \pi(\, \theta \, \lvert \, Y  \,)$  
                    \State \text{Calculate} $X_\theta^{i}(t) \gets$   \text{Solve the forward map}
                \EndFor
                    \State \text{In a point time $t^*$ calculate a credibility region:} 
                    \State $\left( X_\theta^{\alpha/2}(\, t^* \,), \, X_\theta^{1-\alpha/2}(\, t^* \,) \,\right) \gets P \left(\, X_\theta(t^*) \in \left( X_\theta^{\alpha/2}(\, t^* \,), \, X_\theta^{1-\alpha/2}(\, t^* \,) \,\right) \,\right) \geq 1- \alpha$ 
			\end{algorithmic}
		\end{algorithm}

Thus, we can observe that the previous algorithm returns a credible interval that offers a probability $(1 -\alpha)100\%$ of containing the true value of the state variable $X_\theta$ at point time $t^*$. In other words, this method does not allow for quantifying the uncertainty of the entire trajectories of the state variables over the analysis time period, but rather through a discretization of the period. This leads to an error propagation in the uncertainty quantification process.

\subsection{Functional Data Analysis Tool Considered: Modified Band Depth Method}

While the method presented by \cite{Molina2025} quantifies uncertainty on a pointwise basis, our work aims to determine a global credibility band for the estimated solution of the differential equation. To achieve this, we leverage Functional Data Analysis (FDA) frameworks, specifically the Modified Band Depth method (MBD) introduced by \cite{LopezPintado2009}. This method induces an ordering of the curves based on depth, allowing for the derivation of functional quantiles that ultimately define the confidence band. To induce an ordering on the functional sample, we adopt a depth-based approach. This involves measuring the proportion of the domain where a specific curve resides within the ribbons formed by pairs of other curves in the sample. 

Formally, let $X_{i_1}, X_{i_2}$ be two functions defined on a compact domain $I$. The band delimited by these functions is defined as the subset of the plane,
\[
B(X_{i_1}, \, X_{i_2}) = \left\{ \, (t, X) : t \in I, \, \min\{ \, X_{i_1}(t), \, X_{i_2}(t)\} \le X \le \max\{X_{i_1}(t), \, X_{i_2}(t) \, \} \, \right\}.
\]

Let $\lambda$ denote the Lebesgue measure on $I$. We define the normalized Lebesgue measure, $\lambda_{r}$, for any subset $A \subseteq I$ as $\lambda_{r}(A)=\lambda(A) \, / \, \lambda(I).$ Consequently, for a stochastic process $\mathcal{X}$ and a fixed curve $X$, the population Modified Band Depth (MBD) is given by the expected proportion of time that $X$ is contained within the random band formed by two independent copies of $\mathcal{X}$,

\begin{equation}
\text{MBD}(X) = \mathbb{E}\left[ \; \lambda_{r} \left( \, \{t \in I : \, (t, \, X(t)) \, \in B(X_{i_1}(t), \,X_{i_2}(t))\} \, \right) \; \right],
\end{equation}

where $X_{i_1}$ and $X_{i_2}$ are any two independent realizations of $\mathcal{X}$.

Given a sample of independent curves $\{X_1, \ldots, X_N\}$, the sample estimator, $\text{MBD}_N (\, X \,)$, is calculated by averaging over all possible pairs,
\begin{equation}
\text{MBD}_N(X) = \binom{N}{2}^{-1} \sum_{1 \le i_1 < i_2 \le N} \lambda_{r}\left( \, \{t \in I : (t, \, X(t)) \in B(X_{i_1}, X_{i_2})\} \, \right).
\end{equation}

Based on these depth values, we establish a center-outward ordering of the sample, denoted as:
\[
X_{[1]} \prec X_{[2]} \prec \cdots \prec X_{[N]},
\]
where the indices satisfy $\text{MBD}_N(X_{[1]}) \ge \text{MBD}_N(X_{[2]}) \ge \cdots \ge \text{MBD}_N(X_{[N]})$. Here, $X_{[1]}$ represents the deepest curve (sample functional median), while $X_{[N]}$ corresponds to the most outlying observation.

Finally, we construct the $(1-\alpha)$ global credibility band. This region is defined by the envelope of the $\lceil (1-\alpha)N \rceil$ deepest curves. Formally:
\begin{equation}
C_{1-\alpha} = \left\{ (t, Y) : \, L_{\alpha}(t) \le Y \le U_{\alpha}(t), \, \forall t \in I \right\},
\end{equation}
where the lower and upper bounds are given by:
\[
L_{\alpha}(t) = \min_{r=1, \ldots, k} X_{[r]}(t), \quad U_{\alpha}(t) = \max_{r=1, \ldots, k} X_{[r]}(t),
\]
with $k = \lceil (1-\alpha)N \rceil$.

\subsection{Proposed algorithm for uncertainty quantification of state variables} \label{MBD}

Below we present the details of the methodology we propose for uncertainty quantification of state variables in the context of inverse problems, combining tools of Bayesian inference and functional data analysis, to determine credible regions for the trajectories of the state variables throughout the time horizon of interest. Starting from sampled $\theta$ values from its posterior distribution, we calculate multiple trajectories of the state variables, these trajectories are ordered based on their respective estimated Modified Band Depth, and finally a credibility region is calculated for the trajectory of the state variables over the entire analysis time horizon. The following algorithm describes the details of our methodology:

\begin{algorithm}[H] 
			\caption{Proposed algorithm for UQ of State Variables}
            \label{alg_prop}
			\begin{algorithmic}[1]
				\Require \text{Data} $Y=(y_1, \dots , y_n)$, 
                \Require \text{right hand side of DE system} $F(\, \cdot \,)$, 
                \Require \text{prior distribution} $\pi(\cdot)$, 
                \Require likelihood function $f(y \, \lvert \, \theta )$   
			\For{$i \gets 1$ to $N$}
				    \State \text{Sample} $\theta_i \sim \pi(\, \theta \, \lvert \, Y  \,)$  
                    \State \text{Calculate} $X_\theta^{i}(t) \gets$   \text{Solve the forward map}
                \EndFor
                    \State \text{Estimate the MBD of each curve $X_\theta^i$:}
                    \State $\displaystyle{ \text{MBD}_{N}(X_\theta) \gets \binom{N}{2}^{-1} \sum_{1 \le i < j  \le N} \lambda_{r}\left( \, \{t \in I : (t, \, X_\theta(t)) \in B(X_\theta^{i}, X_\theta^{j})\} \, \right) }$
                    \State \text{Order the curves:} 
                    \State $X_\theta^{[1]}(t)\prec X_\theta^{[2]}(t)\prec\cdots \prec X_\theta^{[N]}(t) \; \gets \; X_\theta^{k}(t) \prec X_\theta^{l}(t) \; \; \text{iff} \;\; \text{MBD}_{N}(X_\theta^{k}) \ge \text{MBD}_{N}(X_\theta^{l})$ 
                    \State Calculate the credible region for $X_\theta(t)$:
                    \State $\displaystyle{C_{1-\alpha} \gets \left\{ (\, t, \, X_\theta(t) \,) : \min_{r=1, \ldots, \lceil (1-\alpha)N \rceil} X_\theta^{[r]}(t) \le X_\theta(t) \le \max_{r=1, \ldots, \lceil (1-\alpha)N \rceil} X_\theta^{[r]}(t) \right\} }$
			\end{algorithmic}
		\end{algorithm}

It is important to mention that in the previous algorithm, as the MBD estimator proposed by \cite{LopezPintado2009} is defined for a sample of independent curves, then the quantities $\{ \theta_1, \theta_2, \cdots, \theta_N \}$ must be independent of each other. To guarantee this, must be considered only values generated through the Hamiltonian Monte Carlo algorithm that are at a distance greater than or equal to the integrated autocorrelation time (IAT, for details of the IAT calculation see \cite{molina2022}).

\section{Procedure for methodology validation}

In this section, we present a validation of the methodology that we propose for uncertainty quantification of state variables in the context of inverse problems, through a simulation study associated with the logistic growth model. In the following section, we will present the application of the methodology to a more complex model associated with a neurological phenomenon.

\subsection{Uncertainty Quantification in the Logistic Growth Model}

The logistic growth model describes population growth with limited resources and is defined by the differential equation

$$\frac{dP}{dt} = rP(1-\frac{P}{K}),$$

where $P(t)$ represents the population size at time $t$, $r$ is the intrinsic growth rate, and $K$ is the carrying capacity or maximum sustainable population. The analytical solution to this equation is

$$P(t) = \frac{KP_0}{P_0 + (K-P_0)e^{-rt}} ,$$

where $P_0 = P(0)$ is the initial population size.\\

For our Bayesian inference framework, we assume noisy observations of the population $Y = (y_1, \ldots, y_n)$ taken at times $t = (t_1, \ldots, t_n)$, with the following structure, 

$$y_i = P(\theta, \, t_i) + \varepsilon_i, \quad i = 1,\ldots,n.$$

Where $P$ represents the solution of the logistic growth model for a given parameter set $\theta = (\, r, \, K \,)$, and $\varepsilon_i$ represents measurement errors assumed to be independent and normally distributed with $\varepsilon_i \sim \mathcal{N}(0, \sigma^2)$. We then have that the logistic growth model has only two parameters and an analytical solution, making it ideal for validating the methodology we propose for uncertainty quantification before tackling a more complex neuronal model.\\ 

Under our Gaussian error assumptions, the likelihood function takes the form

$$f(\, Y \, \lvert \, \theta \,) = (2\pi\sigma^2)^{-n/2} \exp\left[-\frac{1}{2\sigma^2} \sum_{i=1}^n(\, y_i - P(\theta, \, t_i) \,)^2\right].$$

For the prior distributions, we considered

$$r \sim \mathcal{N}(\mu_{r}, (0.40\mu_{r})^2),$$
$$K \sim \mathcal{N}(\mu_{K}, (0.40\mu_{K})^2).$$

These priors reflect typical ranges for population growth parameters while remaining sufficiently diffuse to allow the data to influence the posterior distributions. 

Assuming independence between parameters, we get the joint prior distribution is
$$\pi(\theta) = \pi(r) \, \pi(K).$$

By Bayes' theorem, for the posterior distribution we have

$$\pi(\, \theta \, \lvert \, Y \,) \propto f(\, Y \, \lvert \, \theta\,) \; \pi(\theta).$$

To generate values of the posterior distribution we used a variant of HMC through the Python package PyMC3, the No U-Turn Sampler (NUTS) \citep{hoffman2014}. We generated four independent chains with 5000 iterations each, to guarantee: reaching the stationary state, that is, that the distribution of interest was effectively simulated; And based on the criterion proposed by \citet{molina2022}, there is a minimum precision of 3 significant figures in the estimation of parameters.

We adopt a functional data analysis framework that treats the entire population trajectory as a continuous curve rather than discrete time points. In our context, each solution of the logistic growth model  $P(\, \theta, \, t \,)$ represents a continuous function of time parametrized by $\theta$. The posterior distribution over $\theta$ induces a distribution over these functions, creating a random field of population trajectories. \\

To quantify uncertainty in this functional space, we employ a depth-based method for functional data, the Modified Band Depth measure (see section \ref{MBD} for details), defined for the function $P(t)$. Given a sample of functions $\{P_1(t),\ldots,P_N(t)\}$, the estimator of the Modified Band Depth is calculated as,

\begin{equation*}
\text{MBD}_N(\, P \,) = \binom{N}{2}^{-1} \sum_{1 \le i < j \le N} \lambda_{r}\left( \, \{t \in I : (t, \, P(t)) \in B(P_i, \, P_j)\} \, \right).
\end{equation*}

Based on these depth values, we establish a center-outward ordering of the sample, denoted as:
\[
P_{[1]} \prec P_{[2]} \prec \cdots \prec P_{[N]},
\]
where the indices satisfy $\text{MBD}_N(P_{[1]}) \ge \text{MBD}_N(P_{[2]}) \ge \cdots \ge \text{MBD}_N(P_{[N]})$. Here, $P_{[1]}$ represents the deepest curve (sample functional median), while $P_{[N]}$ corresponds to the most outlying observation.

Finally, we construct the $(1-\alpha)$ global credibility band. This region is defined by the envelope of the $\lceil (1-\alpha)N \rceil$ deepest curves, 

\begin{equation}
C_{1-\alpha} = \left\{ (t, W) : \, L_{\alpha}(t) \le W \le U_{\alpha}(t), \, \forall t \in I \right\},
\end{equation}
where the lower and upper bounds are given by:
\[
L_{\alpha}(t) = \min_{r=1, \ldots, k} P_{[r]}(t), \quad U_{\alpha}(t) = \max_{r=1, \ldots, k} P_{[r]}(t),
\]
with $k = \lceil (1-\alpha)N \rceil$.\\ 

\begin{figure}[H]
\centering
\begin{subfigure}{0.6\textwidth}
    \includegraphics[width=\textwidth]{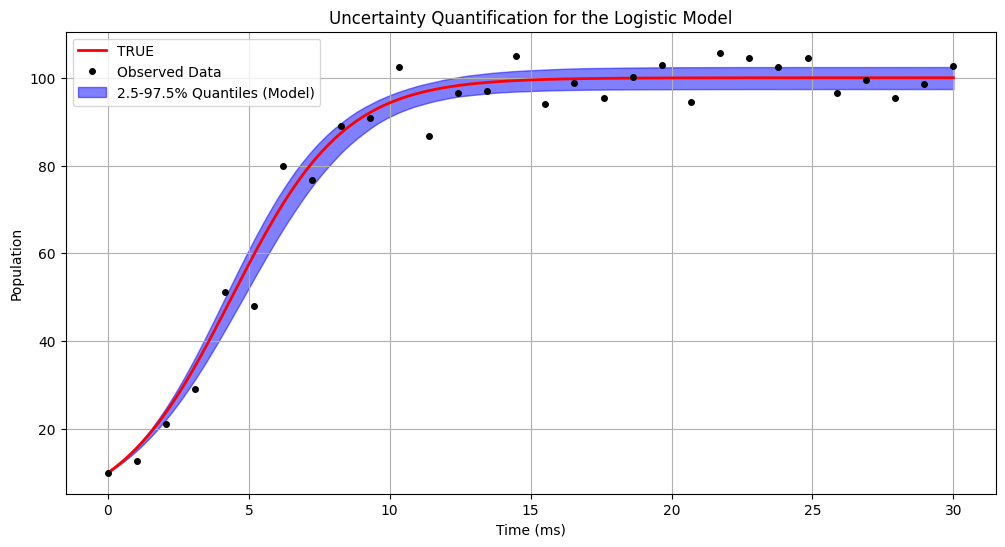}
    \caption{Traditional pointwise uncertainty quantification}
    \label{fig:uncertainty_traditional_logistic}
\end{subfigure}

\vspace{2em} 

\begin{subfigure}{0.6\textwidth}
    \includegraphics[width=\textwidth]{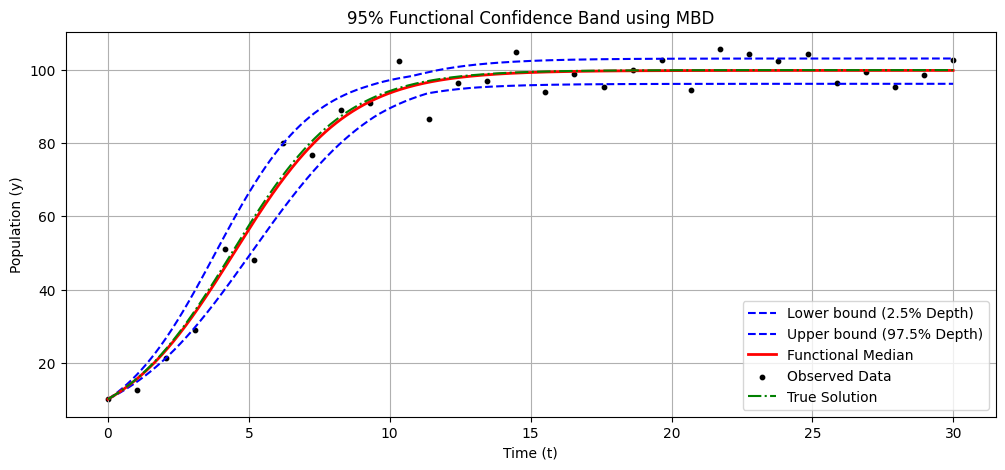}
    \caption{Proposed methodology uncertainty quantification}
    \label{fig:uncertainty_fda_logistic}
\end{subfigure}
\caption{Comparison of the results of uncertainty quantification for the state variable with the proposed methodology and the pointwise method in the context of the logistic growth model.}
\label{fig:uncertainty_logistic}
\end{figure}

Figure \ref{fig:uncertainty_logistic} compares the traditional pointwise uncertainty quantification method proposed by \cite{Molina2025} with the methodology that we propose in this article and that we described in the algorithm \ref{alg_prop}, in the context of the logistic growth model. As seen in Figure \ref{fig:uncertainty_logistic}, the methodology we propose generates a more cohesive representation of the uncertainty, capturing the trajectory's shape and variability across time as a functional entity. In contrast, the traditional pointwise method constructs intervals independently at each time point, potentially missing important temporal dependencies.

 With the aim of making a more extensive comparison of the two methodologies for uncertainty quantification of the state variables, we performed 1000 independent simulations of the logistic growth model, under the conditions already described. The table \ref{tab:coverage_comparison} summarizes the coverage results from the simulations. The table reveals that across all independent batches, our proposal achieved higher coverage rates (averaging 96.40\%) compared to the pointwise method (averaging 80.00\%). The significant performance gap of over 16 percentage points demonstrates our proposal has superior ability to capture the dynamic nature of trajectory uncertainties. This improvement in coverage highlights why the functional approach is more effective for uncertainty quantification in dynamic systems.

\begin{table}[H]
\centering
\caption{Coverage comparison between our proposal and pointwise uncertainty quantification methods}
\label{tab:coverage_comparison}
\begin{tabular}{ccccc}
\hline
1000 & Our proposal & Pointwise & Both Methods & Neither Method \\
Simulations & Coverage (\%) & Coverage (\%) & Coverage (\%) & Coverage (\%) \\
\hline
Average & 96.40 & 80.00 & 80.00 & 3.60 \\
\hline
\end{tabular}
\end{table}

\section{Application: Uncertainty Quantification for the Hodgkin-Huxley Model}

The Hodgkin-Huxley model, proposed by \cite{hh-model}, describes the electrical behavior of excitable cells using a system of nonlinear ordinary differential equations. The model represents the cell membrane as a capacitor with voltage-dependent conductances for sodium (Na$^+$), potassium (K$^+$), and a leakage current. The main equation governing the membrane potential $V$ is

\[ C_m\frac{dV}{dt} = -(I_{Na} + I_K + I_l), \]

where $C_m$ is the membrane capacitance, and $I_{Na}$, $I_K$, and $I_l$ are the currents through the sodium, potassium, and leakage channels, respectively. The leakage current represents the passive flow of ions through non-gated channels. These currents are described by

\begin{align*}
	I_{Na} &= g_{Na}m^3h(V-V_{Na}), \\
	I_K &= g_Kn^4(V-V_K), \\
	I_l &= g_l(V-V_l).
\end{align*}

The parameter estimation focuses on several key components of the Hodgkin-Huxley model. The maximum conductances ($g_{Na}$, $g_K$, $g_l$) represent the membrane's peak permeability to specific ions when all channels of that type are open. The reversal potentials ($V_{Na}$, $V_K$, $V_l$) characterize the equilibrium points where net ionic currents are zero, determined by concentration gradients across the membrane. The membrane capacitance ($C_m$) influences the cell's response time to current changes. Each parameter plays a distinct role in shaping the action potential: sodium conductance primarily controls the depolarization rate, potassium conductance governs repolarization dynamics, and membrane capacitance modulates the overall temporal characteristics of the response.

The variables \(m\), \(h\), and \(n\) are dimensionless quantities that range from 0 to 1, such that the variable \(m\) is associated with the activation of the sodium (Na) channel, \(h\) with its inactivation, and \(n\) with the activation of the potassium (K) channel. These variables evolve according to the following equations

\begin{align*}
	\frac{dm}{dt} &= \alpha_m(V)(1-m) - \beta_m(V)m, \\
	\frac{dh}{dt} &= \alpha_h(V)(1-h) - \beta_h(V)h, \\
	\frac{dn}{dt} &= \alpha_n(V)(1-n) - \beta_n(V)n.
\end{align*}

The $\alpha_i(V)$ and $\beta_i(V)$ functions are voltage-dependent rate constants that describe the transition rates between open and closed states for each gating variable. The original Hodgkin-Huxley formulation used empirical fits to experimental data for these functions

\begin{align*}
	\alpha_n(V) &= 0.01 \frac{10 - V}{\exp\left(\frac{10 - V}{10}\right) - 1}, & \beta_n(V) &= 0.125 \exp\left(-\frac{V}{80}\right), \\[3pt]
	\alpha_m(V) &= 0.1 \frac{25 - V}{\exp\left(\frac{25 - V}{10}\right) - 1}, & \beta_m(V) &= 4 \exp\left(-\frac{V}{18}\right), \\[3pt]
	\alpha_h(V) &= 0.07 \exp\left(-\frac{V}{20}\right), & \beta_h(V) &= \frac{1}{\exp\left(\frac{30 - V}{10}\right) + 1}.
\end{align*}

The estimation of Hodgkin-Huxley model parameters represents an inverse problem in computational neuroscience. Given membrane potential observations $V = (V_1, \dots, V_n)$ at times $t = (t_1, \dots, t_n)$, we aim to recover the underlying model parameters $\theta = (g_{Na}, g_K, g_l, V_{Na}, V_K, V_l, C_m)$. The relationship between observations and parameters follows an additive noise model

\begin{equation*}
	V_i = V_h(\,\theta, \, t_i \,) + \varepsilon_i,
\end{equation*}

where $V_h(\, \theta, \, t_i \,)$ represents the Hodgkin-Huxley model solution at time $t_i$ for parameters $\theta$ and $\varepsilon_i$ represent measurement error with a specific distribution. This formulation presents several challenges: the nonlinearity of the Hodgkin-Huxley model makes the parameter-to-observation map highly complex, evaluating this map requires numerical solution of a stiff system of ODEs, the high dimensionality of the parameter space complicates the estimation process, and measurement noise introduces additional uncertainty into the estimation process. These characteristics make the Hodgkin-Huxley inverse problem particularly challenging and necessitate advanced estimation techniques.\\

Given measurements of the membrane potential $V = (V_1, \ldots, V_n)$ taken at times $t = (t_1, \ldots, t_n)$, we assume the following observation model

$$V_i = V_h(\,\theta, \, t_i \,) + \varepsilon_i, \quad i = 1,\ldots,n.$$

Where $V_h$ represents the solution of the Hodgkin-Huxley equations for a given parameter set $\theta = (g_{Na}, g_K, g_L, V_{Na}, V_K, V_L, C_m)$, and $\varepsilon_i$ represents measurement errors assumed to be independent and normally distributed with $\varepsilon_i \sim \mathcal{N}(0, \sigma^2)$.

Under our Gaussian error assumptions, the likelihood function takes the form

$$f(\, V \,\lvert \, \theta \,) = (2\pi\sigma^2)^{-n/2} \exp\left[-\frac{1}{2\sigma^2} \sum_{i=1}^n(\, V_i - V_h(\,\theta, \, t_i \,) \,)^2\right].$$

For the prior distributions $\pi(\theta)$, we consider both physical constraints and existing knowledge about parameter ranges. For the conductances and membrane capacitance, which must be strictly positive due to their physical nature, we employ log-normal distributions, where $$\log(g_{Na}) \sim \mathcal{N}(\mu_{g_{Na}}, (0.25\mu_{g_{Na}})^2),$$
$$\log(g_K) \sim \mathcal{N}(\mu_{g_K}, (0.25\mu_{g_K})^2),$$
$$\log(g_L) \sim \mathcal{N}(\mu_{g_L}, (0.25\mu_{g_L})^2),$$ 
$$\log(C_m) \sim \mathcal{N}(\mu_{C_m}, (0.25\mu_{C_m})^2).$$

Here, each $\mu$ parameter represents the logarithm of the nominal value for that parameter, with the standard deviation set to 25\% of the mean value to provide substantial variability while maintaining physiologically plausible ranges.\\

For the reversal potentials, which can be either positive or negative, we use normal distributions centered at their nominal values,
$$V_{Na} \sim \mathcal{N}(\mu_{V_{Na}}, (0.25 \lvert \mu_{V_{Na}} \lvert)^2),$$
$$V_K \sim \mathcal{N}(\mu_{V_K}, (0.25 \lvert \mu_{V_K} \lvert)^2),$$
$$V_L \sim \mathcal{N}(\mu_{V_L}, (0.25 \lvert \mu_{V_L} \lvert)^2).$$

The prior distributions are chosen to be relatively diffuse, with standard deviations set again to 25\% of the absolute value of their means, ensuring the inference problem remains challenging while maintaining physiologically plausible ranges.\\

Assuming independence between parameters, we get the joint prior distribution is
$$\pi(\theta) = \pi(g_{Na}) \;  \pi(g_K) \; \pi(g_L) \; \pi(V_{Na}) \; \pi(V_K) \; \pi(V_L) \; \pi(C_m).$$

By Bayes' theorem, for the posterior distribution we have

$$\pi(\, \theta \, \lvert \, V \,) \propto f(\, V \, \lvert \, \theta\,) \; \pi(\theta).$$

To generate values of the posterior distribution we used a variant of HMC through the Python package PyMC3, the No U-Turn Sampler (NUTS) \citep{hoffman2014}. We generated four independent chains with 5000 iterations each, to guarantee: reaching the stationary state, that is, that the distribution of interest was effectively simulated; And based on the criterion proposed by \citet{molina2022}, there is a minimum precision of 3 significant figures in the estimation of parameters.

Table \ref{tab:parameters} presents the results of this Bayesian inference process, showing the true parameter values used to generate the synthetic data, alongside the posterior means and standard deviations estimated from our sampling procedure. These results validate our Bayesian approach for reliable parameter inference, as the parameter estimates closely recover the true values used to generate the data, with small standard deviations indicating precise inference.

\begin{table}[h!]
	\centering
	\begin{tabular}{lccc}
		\hline
		Parameter & True Value & Posterior Mean & Standard Deviation \\
		\hline
		$g_{Na}$ ($\mu$S) & 120.0 & 116.576 & 7.068 \\
		$g_K$ ($\mu$S) & 36.0 & 35.295 & 1.879 \\
		$g_l$ ($\mu$S) & 0.3 & 0.291 & 0.057 \\
		$V_{Na}$ (mV) & 115.0 & 114.305 & 1.571 \\
		$V_K$ (mV) & -12.0 & -11.894 & 0.633 \\
		$V_l$ (mV) & 10.6 & 10.674 & 2.664 \\
		$C_m$ ($\mu$F) & 1.0 & 1.005 & 0.041 \\
		\hline
	\end{tabular}
	\caption{Parameter estimates from Bayesian inference}
	\label{tab:parameters}
\end{table}

While traditional parameter estimation methods yield point estimates or marginal credible intervals, they fail to capture the functional nature of the neuronal dynamics. We adopt a functional data analysis framework that treats the entire voltage trajectory as a continuous curve rather than discrete time points. This approach offers several advantages for uncertainty quantification in the Hodgkin-Huxley model. Functional data analysis considers data as observations of continuous functions rather than multivariate vectors \citep{Ramsay2005FDA}. In our context, each solution of the Hodgkin-Huxley model $V_h(\, \theta, \, t \,)$ represents a continuous function of time parametrized by $\theta$. The posterior distribution over $\theta$ induces a distribution over these functions, creating a random field of potential voltage trajectories. \\

To quantify uncertainty in this functional space, we employ a depth-based method for functional data, the Modified Band Depth measure (see section \ref{MBD} for details), defined for the function $V(t)$. Given a sample of functions $\{V_1(t),\ldots,V_N(t)\}$, the estimator of the Modified Band Depth is calculated as,

\begin{equation*}
	\text{MBD}_N(\, V \,) = \binom{N}{2}^{-1} \sum_{1 \le i < j \le N} \lambda_{r}\left( \, \{t \in I : (t, \, V(t)) \in B(V_i, \, V_j)\} \, \right).
\end{equation*}

Based on these depth values, we establish a center-outward ordering of the sample, denoted as:
\[
V_{[1]} \prec V_{[2]} \prec \cdots \prec V_{[N]},
\]
where the indices satisfy $\text{MBD}_N(V_{[1]}) \ge \text{MBD}_N(V_{[2]}) \ge \cdots \ge \text{MBD}_N(V_{[N]})$. Here, $V_{[1]}$ represents the deepest curve (sample functional median), while $V_{[N]}$ corresponds to the most outlying observation.

Finally, we construct the $(1-\alpha)$ global credibility band. This region is defined by the envelope of the $\lceil (1-\alpha)N \rceil$ deepest curves, 

\begin{equation}
	C_{1-\alpha} = \left\{ (t, W) : \, L_{\alpha}(t) \le W \le U_{\alpha}(t), \, \forall t \in I \right\},
\end{equation}
where the lower and upper bounds are given by:
\[
L_{\alpha}(t) = \min_{r=1, \ldots, k} V_{[r]}(t), \quad U_{\alpha}(t) = \max_{r=1, \ldots, k} V_{[r]}(t),
\]
with $k = \lceil (1-\alpha)N \rceil$. The region $C_{1-\alpha}$ provides a comprehensive characterization of uncertainty that respects the functional nature of the voltage dynamics, capturing complex temporal correlations that traditional pointwise intervals ignore. The integration of functional data analysis with Bayesian inference offers a powerful framework for uncertainty quantification in the Hodgkin-Huxley model. By representing uncertainty as regions in function space rather than parameter space, our approach preserves the temporal structure of action potentials and provides neuroscientists with more intuitive and accurate uncertainty representations.\\

\begin{figure}[H]
	\centering
	\begin{subfigure}{0.75\textwidth}
		\centering
		\includegraphics[width=\textwidth]{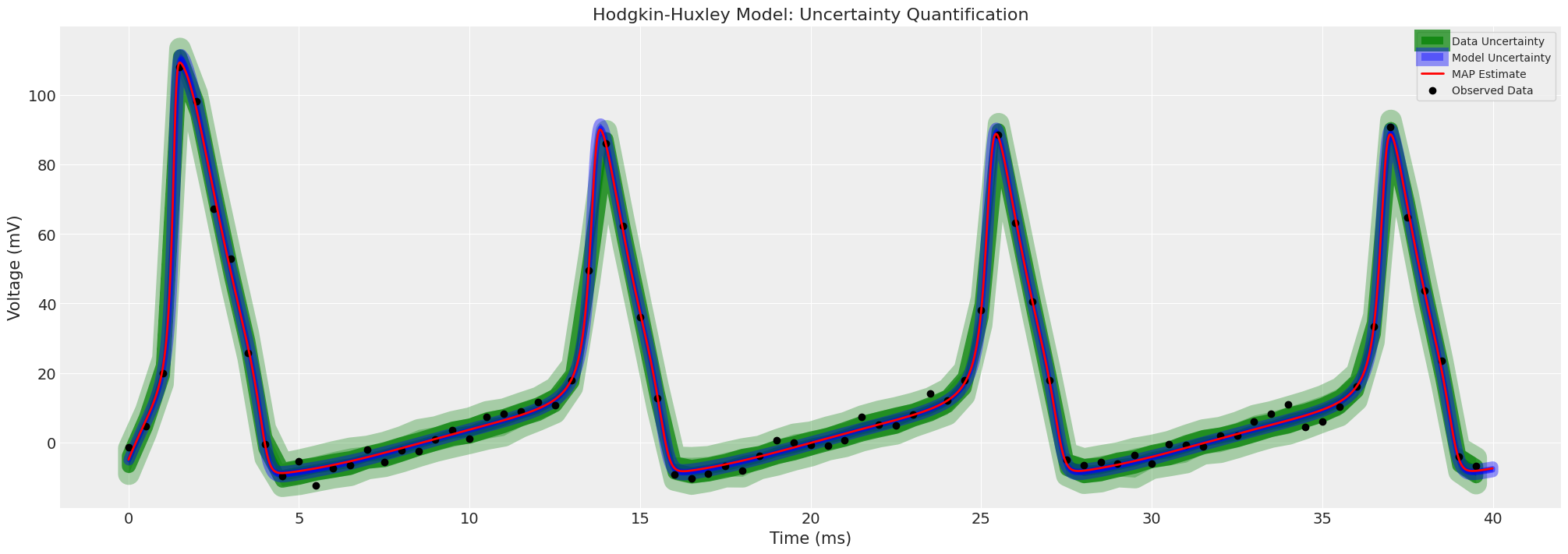}
		\caption{Traditional pointwise uncertainty quantification}
		\label{fig:uncertainty_traditional_hh}
	\end{subfigure}
	
	\vspace{2em} 
	
	\begin{subfigure}{0.75\textwidth}
		\centering
		\includegraphics[width=\textwidth]{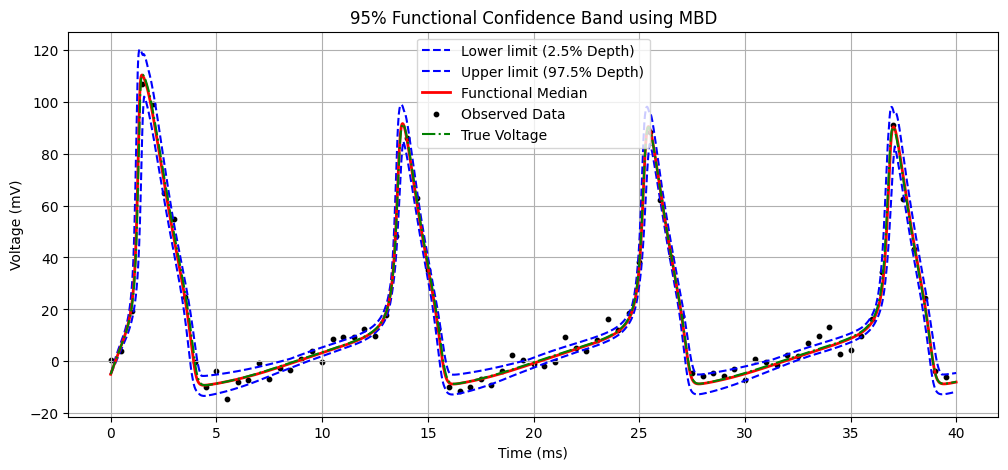}
		\caption{Functional data analysis uncertainty quantification}
		\label{fig:uncertainty_fda_hh}
	\end{subfigure}
	
	\caption{Comparison of the results of uncertainty quantification for the state variable with the
		proposed methodology and the pointwise method in the context of the Hodgkin-Huxley model}
	\label{fig:uncertainty_hh}
\end{figure}

Building on the validation results from the logistic growth model, we applied both traditional and functional uncertainty quantification methods to the Hodgkin-Huxley model. Figure \ref{fig:uncertainty_hh} presents the results for the Hodgkin-Huxley model. The traditional pointwise method (Figure \ref{fig:uncertainty_traditional_hh}) produces credible intervals that, while capturing the general magnitude of uncertainty, fail to account for the functional nature of the voltage traces. This leads to fragmented credible regions that do not adequately represent the temporal correlations inherent in action potential dynamics. In contrast, our proposal (Figure \ref{fig:uncertainty_fda_hh}) generates a coherent credible region that respects the functional characteristics of the voltage trajectories. The MBD-based regions capture the uncertainty throughout the entire action potential, including the rapid depolarization phase, repolarization, and hyperpolarization. These results demonstrate that our proposal provides a more comprehensive and physiologically meaningful quantification of uncertainty in the Hodgkin-Huxley model, capturing the functional variability of action potentials while respecting their temporal structure.

\section{Conclusions}

Currently, the literature offers very few alternatives for analyze the propagation of uncertainty of state variables, in the context of inverse problems. The existing options are limited to constructing pseudo-credible regions or quantifying their uncertainty at isolated points. The pseudo-credible regions are determined by plot multiple feasible trajectories of the state variables, and observing the region they occupy. An example of a methodology for uncertainty quantification of state variables at isolated points, is that proposal by \cite{Molina2025}, which involves generating credible regions for each point in time $t^*$ within the analysis period, based on the values of $\theta$, obtained from its posterior distribution. However this method does not allow for quantifying the uncertainty of the entire trajectories of the state variables over the analysis time period, but rather through a discretization of the period. This leads to an error propagation in the uncertainty quantification process.\\ 

In this article, we propose a methodology that
combines tools of Bayesian inference, with Hamiltonian Monte Carlo sampling employed for
efficient posterior exploration, and functional data analysis, specifically the Modified Band Depth method, to determine credible regions for the trajectories of the state variables throughout the time horizon of interest. Our methodology is computational, and as such, we present an algorithm that allows for its clear and simple implementation. We validated the methodology through a simulation study associated with the logistic growth model, which shows that our proposal captures a higher proportion of state variable trajectories than traditional pointwise analysis methods, our proposal generated credible regions that contained the true trajectory of the state variables 96.4\% of the times, versus 80\% that the regions of the pointwise method did.\\

Furthermore, we expose an 
application of our methodology to a non-trivial model
associated with a neurological phenomenon, the Hodgkin-Huxley model that describes the electrical behavior of excitable cells. In this application,  our methodology effectively captured rapid voltage changes during action potentials while respecting biophysical constraints. This confirms the idea that the methodology we propose effectively captures the
time-dependent dynamics of the state variables.\\

Possible future research related to the work presented in this article includes extending the proposed methodology to inverse problems where the evolution of state variables is analyzed not only over time, but simultaneously in time and space. This is common in phenomena represented by partial differential equations. And, specifically in the context of applications to neurological phenomena, it would be worthwhile to explore the use of our methodology for the characterization and diagnosis of diseases.


\bibliographystyle{newapa}
\addcontentsline{toc}{chapter}{References}
\bibliography{references}

\end{document}